\documentclass[useAMS,usenatbib,usegraphicx]{mn2e}

\usepackage{graphicx,times}

\title[The sudden appearance of CO emission in LHA~115-S~65]{The sudden appearance of CO emission in LHA~115-S~65
\thanks{Based on observations 
obtained with ESO telescopes at the La Silla Paranal Observatory under program ID 088.D-044, and
 at the Gemini Observatory, which is operated by the Association of Universities for
    Research in Astronomy, Inc., under a cooperative agreement with the
    NSF on behalf of the Gemini partnership: the National Science
    Foundation (United States), the Science and Technology Facilities
    Council (United Kingdom), the National Research Council (Canada),
    CONICYT (Chile), the Australian Research Council (Australia),
    Minist\'{e}rio da Ci\^{e}ncia, Tecnologia e Inova\c{c}\~{a}o (Brazil)
    and Ministerio de Ciencia, Tecnolog\'{i}a e Innovaci\'{o}n Productiva
    (Argentina), under program ID GS-2010B-Q-31.
}}
\author[M.E. Oksala et al.]{M.E.~Oksala,$^{1}$\thanks{E-mail: 
oksala@sunstel.asu.cas.cz 
} 
M.~Kraus,$^{1}$  M.L.~Arias,$^{2,3}$ M.~Borges Fernandes,$^{4}$
\newauthor
L.~Cidale,$^{2,3}$ M.F.~Muratore,$^{2,3}$ M.~Cur\'e$^{5}$\\
$^{1}$ Astronomick\'y \'ustav, Akademie v\v{e}d \v{C}esk\'e republiky,
Fri\v{c}ova 298, 251\,65 Ond\v{r}ejov, Czech Republic\\ 
$^{2}$ Departamento de Espectroscop\'ia Estelar, Facultad de Ciencias
Astron\'omicas y Geof\'isicas, Universidad Nacional de La Plata,\\
~~~Paseo del Bosque s/n, B1900FWA, La Plata, Argentina\\
$^{3}$ Instituto de Astrof\'isica de La Plata, CCT La Plata, CONICET-UNLP,
Paseo del Bosque s/n, B1900FWA, La Plata, Argentina\\
$^{4}$ Observat\'orio Nacional, Rua General Jos\'e Cristino 77,
20921-400 S\~ao Cristov\~ao, Rio de Janeiro, Brazil\\
$^{5}$ Departmento de F\'isica y Astronom\'ia, Facultad de Ciencias, 
Universidad de Valpara\'iso Av. Gran Breta\~na 1111, Casilla 5030, Valpara\'iso, Chile
}

\begin{document}

\date{Accepted 2012 July 25. Received 2012 July 25; in original form 2012 June 12}

\pagerange{\pageref{firstpage}--\pageref{lastpage}} \pubyear{2012}

\maketitle

\label{firstpage}

\begin{abstract}

Molecular emission has been detected in several Magellanic Cloud B[e] 
supergiants.  In this Letter, we report on the detection of CO band 
head emission in the B[e] supergiant LHA 115-S 65, and present a
 $K$-band near-infrared spectrum obtained with the Spectrograph for INtegral Field 
Observation in the Near-Infrared (SINFONI; R=4500) on the ESO VLT UT4 telescope. 
The observed molecular band head emission in S\,65 is quite surprising in light of a previous non-detection
by \citet{McGregor89}, as well as a high resolution (R=50000) Gemini/Phoenix spectrum of this star taken 
nine months earlier showing no emission.  Based on analysis of the optical spectrum by
\citet{Kraus10}, we suspect that the sudden appearance of molecular emission could be due to density 
build up in an outflowing viscous disk, as seen for Be stars.  This new discovery, combined with variability in two other 
similar evolved massive stars, indicates an evolutionary link between B[e] supergiants and LBVs.

\end{abstract}


\section{Introduction}

Massive stars evolving off the main-sequence pass through several phases
of strong mass-loss before ending their lives as supernovae.  
B[e] supergiants (B[e]SG) represent one of these phases.  These stars are surrounded by massive 
disks or rings of cool and dense material, confirmed by polarimetry 
\citep{Magalhaes92,OD99,Melgarejo,Magalhaes06} and optical
interferometry observations \citep{DS07}. They are hence ideal environments 
for efficient dust and molecule condensation.
Dust is evident from a prominent near- and mid-infrared excess emission
\citep[e.g.,][]{Zickgraf86, Bonanos09, Bonanos10} and from spectroscopic features
in the infrared indicating that the disk contains both 
oxygen-rich (crystalline silicates) and carbon-rich (polycyclic aromatic 
hydrocarbons, PAHs) dust \citep{Kastner10}. The presence of molecules in the 
disks of B[e]SGs is obvious from detections of TiO band emission at optical 
wavelengths \citep{Zickgraf89,Torres} and of the much
more prominent CO band emission arising at near-infrared wavelengths 
\citep{McGregor88a,McGregor88b,McGregor89,Morris96}.

The study of CO molecules in the spectra of B[e]SGs is an important tool to understand the inner portion 
of their dusty disks. CO first-overtone band head structure, arising at 2.3 $\mu$m, is
particularly sensitive to the motion of the hot gas disk allowing study of the disk kinematics
and structure (Keplerian disk, slow outflowing winds, or accretion disks).  Complementary
to this molecule, optical forbidden lines, mainly [O~{\sc i}] \citep{Kraus07,Kraus10}
and [Ca~{\sc ii}] lines \citep{Kraus10,Aret}, are excellent tracers of the inner regions 
of the disk.  In addition, the ratio, $^{12}$CO/$^{13}$CO in B[e]SGs could be used as an 
age indicator of the star \citep{Kraus09,Liermann}, allowing constraint on their evolutionary 
phase with respect to other evolved phases of massive stars, i.e. yellow hypergiants (YHGs) 
\citep{Muratore}.

\begin{table*}
\begin{minipage}{140mm}
\caption{Parameters of S\,65.}   
\label{tab:parameters} 
\begin{tabular}{l c c c c c c c}
\hline                 
Object & Sp. Type & $E(B-V)$ & $T_{\rm eff}$ & $L$ & $V_{\rm{sys}}$ & $v \sin i$ & References \\    
 &  & & (10$^3$K) & (10$^5L_{\odot}$) & (km s$^{-1}$) & (km s$^{-1}$) &   \\ 
\hline   
 S\,65  & B2-3    & 0.15-0.20 & 17    & 5.0   & 191  & 150 & (1,2,3) \\
\hline      
\end{tabular}

\medskip
References: (1) \citet{Zickgraf86}; (2) \citet{Zickgraf00}; (3) \citet{Kraus10}.\\

\end{minipage}
\end{table*}

\begin{figure*}
\centering
\includegraphics[width=140mm]{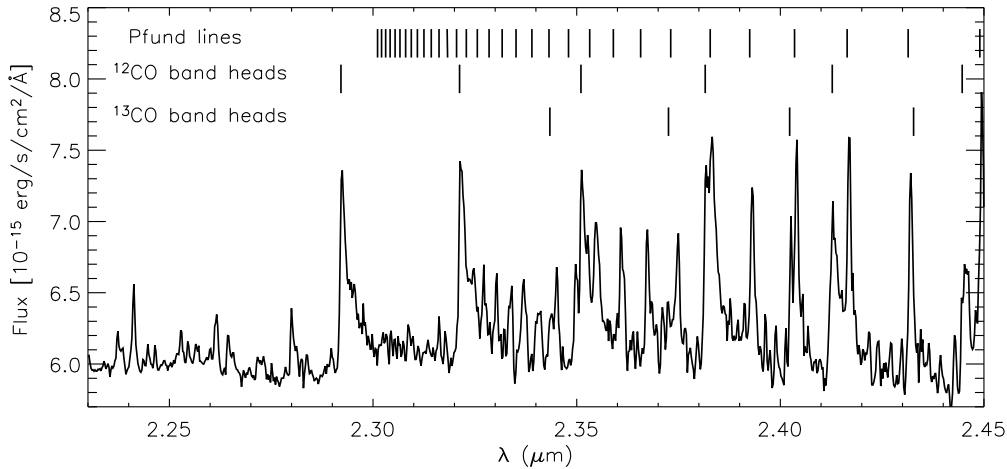}
\caption{Flux calibrated SINFONI CO band spectra of S\,65.
The location of the $^{12}$CO and $^{13}$CO band heads and the lines of the Pfund series are indicated.}
\label{COspec}
\end{figure*}

Recently, we obtained data for a near-infrared survey to study the molecular 
circumstellar material (in terms of CO band emission) and possible 
evolutionary links between YHGs and B[e]SGs based on their 
$^{13}$C footprint \citep{Liermann, Muratore}.  
The survey sample consists of Galactic B[e]SG candidates with CO band emission 
and all Magellanic Cloud (MC) B[e]SGs (except the stars S\,111 from the Large (LMC) and S\,23 from the 
Small Magellanic Cloud (SMC) excluded due to 
magnitude limitations). 
In this Letter, we report on one result from this survey,
the new and sudden detection of CO emission
in the SMC B[e]SG LHA~115-S~65 (hereafter, S\,65).
We discuss the significance of this event in the context of disk kinematics
and stellar evolution.

\citet{Zickgraf86} studied the photometric and spectroscopic
properties of several B[e] supergiants, and determined the stellar parameters
for the SMC star S\,65 (summarized in Table \ref{tab:parameters}).  The observed
optical spectrum consists of many faint absorption lines (e.g., He\,{\sc i}, N\,{\sc ii} and Si\,{\sc ii}),
but also emission lines of Fe\,{\sc ii}, [Fe\,{\sc ii}], and [O\,{\sc i}].  The Fe\,{\sc ii} emission
lines show a central absorption; forbidden emission lines are not double peaked.
Hydrogen Balmer lines have broad wings and consist of both an emission and an absorption component.  
Absorption lines of Ti\,{\sc ii} and Cr\,{\sc ii} are assumed to be shell absorption features.  
Optical spectra of S\,65 taken over the past 30 years do not 
show any variation \citep{Kraus10}; no brightness variations 
exceeding 0.02 mag have been reported since the initial observation 
in 1960 \citep{Zickgraf86}.

\section{Observations}
\label{sec:obs}
We obtained a high-quality, medium resolution (R = 4500) $K$-band (1.95-2.45 $\mu$m) 
spectrum of the SMC B[e] supergiant S\,65 using the Spectrograph for 
INtegral Field Observation in the Near-Infrared 
\citep[SINFONI;][]{Eisenhauer03, Bonnet04} on the VLT UT4 telescope.  
The observation was taken on 2011 October 6 with an 8 $\times$ 8 arcsec$^2$ field of view and an AB 
nod pattern.  A B-type standard star was observed at similar airmass for telluric correction and 
flux calibration.

Data reduction was performed with the SINFONI pipeline (version 2.2.9).  Raw frames were 
corrected for bad pixels, flat fields, distortion, and then wavelength calibrated.  The standard
star observation was similarly reduced.  For flux calibration, 
the standard star spectrum was scaled with the corresponding Kurucz flux model 
\citep{Kurucz93} to its Two Micron All-Sky Survey (2MASS) \citep{Skrutskie06} 
$K_{\rm S}$-band magnitude to create a calibration curve.  The I{\sevensize RAF} task \textit{telluric} 
was used to remove atmospheric lines using the B-type telluric standard spectrum.
The final spectrum has a signal-to-noise ratio S/N $\sim$ 250-300.  The spectrum was corrected for the 
systemic and heliocentric velocities and dereddened with the $E(B-V)$ value 
in Table \ref{tab:parameters} according to the interstellar extinction relation 
of \citet{Howarth83}.  Fig.\,\ref{COspec} shows the final spectrum displaying the CO band heads,
located longwards of $2.29\,\mu$m.

A spectrum of S\,65 was obtained on 2011 January 5 with the Phoenix
high resolution (R=50000) near-infrared spectrograph at the Gemini
Observatory using the K 4396 filter centered at 2.2725
microns.  The observations were reduced using standard I{\sevensize RAF} tasks.  
Observation was taken with
the offset pattern ABBA, and pairs were subtracted to remove sky
background.  The spectrum was flat fielded, telluric-corrected, and wavelength
calibrated.  We again selected a B-type telluric standard.  The final spectrum is 
shown in the top portion of Fig.~\ref{noCO}.

\section{Results}
\label{sec:results}

\subsection{CO detection in S\,65}
\label{S65}

From visual inspection of the S\,65 CO spectrum (Fig.\,\ref{COspec}),
the band head structures appear quite narrow.  Based on the analysis of 
\citet{Kraus10}, the CO should be located beyond 3000 R$_{\star}$ in the edge-on viewed disk,
where the Keplerian rotation has reached a value $< 5$\,km\,s$^{-1}$. 
Therefore, the profiles contain no significant kinematical 
broadening, leaving mainly the instrumental broadening.  The first three band heads appear approximately
equal in strength, while the intensity of the higher band heads quickly drops.  This behavior hints towards a relatively cool CO gas 
(T $<$ 2800\,K), and a moderately high CO column density (0.5-5$\times$10$^{21}$ cm$^{-2}$), which is high enough so that the transitions
at longer wavelengths start to become marginally optically thick.  Such conditions were also recently found for two B[e]SGs 
in the LMC \citep{Liermann}.  
The continuum is flat and Pfund lines contaminate the region where these CO emission 
bands appear.  $^{13}$CO emission is present in the spectrum, however proper modeling 
of the spectrum is necessary to determine the $^{12}$CO$/^{13}$CO ratio.

\begin{figure}
\centering
\includegraphics[width=80mm]{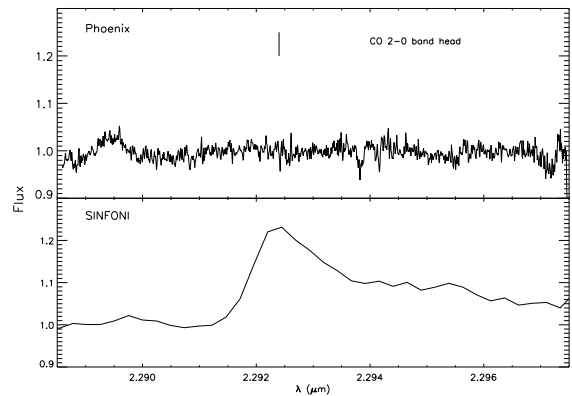}
\caption{Top: Normalized Phoenix high-resolution (R=50000) $K$-band spectrum of S\,65.
Bottom: Normalized SINFONI (R=4500) $K$-band spectrum of S\,65.  
The position of the 2-0 band head of CO is indicated in the top frame.  
Note the absence of detectable emission in the Phoenix spectrum, 
taken just nine months prior to the SINFONI spectrum.  The emission feature at 2.2895 microns 
is likely a blend of two Si lines.}
\label{noCO}
\end{figure}

The discovery of CO band emission in S\,65 is quite surprising, considering the 
previous non-detection by \citet{McGregor89}, further augmented by 
a new high resolution (R=50000) Phoenix spectrum of the star 
(top portion of Fig.\,\ref{noCO}) displaying absolutely no evidence of the 2-0 first 
overtone band head of CO. The absence of CO in this detailed spectrum is 
particularly perplexing considering it was acquired just nine months previous 
to our SINFONI observation.  This very sudden appearance of molecular emission
indicates that the disk of S\,65 must in fact be variable.

\subsection{CO variability of other evolved massive stars}
\label{COvar}

While the abrupt appearance of CO emission in S\,65 is interesting, it is not the first instance of such
variability among evolved massive stars.  LHA~115-S~18 (S\,18), another B[e]SG located in the SMC,
 has shown fluctuation in a number of different spectral features.  In the UV and optical spectra 
observed by \citet{Shore87}, C~{\sc iv} and  N~{\sc iv} resonance lines were variable, as well as 
the He {\sc ii} 4686~\AA~line.  In fact, He~{\sc ii} was absent in the spectrum of \citet{Zickgraf89}.
The spectrum also contained a broad feature at 6159~\AA~attributed to 
TiO emission.  It was therefore expected by \citet{McGregor89} that the $K$-band spectrum of 
S\,18 would show CO emission, as TiO presumably traces a more dense region of material than CO.  
Curiously though, the spectrum of this star showed no 
evidence of CO band head structure.  The star was observed again in the $K$-band region by \citet{Morris96},
who found that the spectrum had not only detectable CO emission, but other spectral changes compared
to the spectrum of \citet{McGregor89}.  The He~{\sc i} 2.058~$\mu$m line strength greatly increased, 
 and the He~{\sc i} 2.112 $\mu$m line (which is typically in absorption for B-type stars) was 
found to be in emission.  While \citet{Shore87} suggested that the changes seen in the spectrum 
of S\,18 were due to interaction between a companion star and the stellar wind, \citet{Morris96} 
determined that the varying He~{\sc i} was not due to the wind, but instead 
a result of a recent increase in density such as a variable equatorial outflow, and likely related to the 
appearance of CO.  \citet{Torres} have recently detected variability in various optical spectral lines, as well as the presence of Raman-scattered
lines that provide evidence of a dense H~{\sc I} region.  A recent SINFONI $K$-band observation confirms the continued presence of the CO emission
in S\,18 (Oksala et al., in preparation).

Although observed spectral variation in Luminous Blue Variables (LBVs) is not noteworthy, 
the presence and variability of first overtone CO band head emission in the spectrum of HR~Car is unique.  
No other LBV has shown such spectral features to date.  
The $K$-band spectrum of HR~Car was first observed by
 \citet{McGregor88a} during three separate periods from 1984 to 1986.  The CO emission was 
clearly detected during the first period, marginally detected during the second, and completely absent 
during the third.  A later observation by \citet{Morris97} revealed the emission had reappeared, 
even stronger than observed previously.  The authors find a photometric brightness correlation with 
the appearance and disappearance of $K$-band CO emission, and determine the location 
of the CO emitting region to be at 3-4~R$_{\star}$.  At this distance, a shielding mechanism is 
necessary to protect the CO molecules from the stellar radiation.  As in S\,65, the most plausible model
for the production of this emission is an outflowing disk.  Although HR~Car hosts a large bipolar nebula, 
spectroscopy \citep{Nota97} and polarization \citep{Clampin95} show support for a more dense, 
flattened central region or waist, likely the location of the CO emitting region.   

\section{Discussion}

\subsection{Disk formation in B[e]SGs}

The kinematics determined within the gaseous disk parts around 
S\,65 were found to be (quasi-)Keplerian by \citet{Kraus10}.  
Furthermore, the authors found that the outflow velocity of the
disk material seems to decrease with distance from the star.
This type of disk has been seen for rapidly-rotating Be stars,
and is likely driven by viscosity \citep[see e.g.,][]{Porter99,Okazaki01,Jones08}.
If the velocity decreases to a static point, the material closer to the star 
will still move outwards, creating an accumulation or high density of material 
at that distance in the disk.  Detectable CO band emission would thus only be observed as soon
as the density is high enough for the band heads to peak out above the continuum.

As in Be stars \citep{Porter99}, the disks of B[e] stars are thought to be formed via 
a two-component stellar wind \citep[see e.g., ][]{Zickgraf85,Zickgraf06} consisting 
of a fast polar wind and a slow equatorial disk wind.
It is possible that the disks of B[e]SGs are similarly outflowing viscous decretion disks,
creating the conditions for detection of molecular emission. 
The stability of the molecular emission in the disks of these highly luminous stars is 
yet to be determined.  Could the observed CO simply be a feature of a transient disk, as in the 
Be stars, or are the circumstellar environments of these massive supergiants more persistent?
Could these disks be products of LBV-type eruption, but on a smaller scale, or are they remnants
of a red supergiant phase?  To determine the mechanism creating and changing these 
disks, we need more consistent monitoring of this whole class of stars 
with data that will allow in depth study of the kinematics, as well as determination of the
evolutionary state.

\subsection{Evolutionary links}

Besides B[e]SGs, the much cooler yellow hypergiants (YHGs) also
display CO band emission.  These stars represent another 
post-main sequence evolutionary phase in which stars undergo strong 
mass-loss \citep[e.g.,][]{deJager} forming a disk, ring or shell 
of high-density material. YHGs are proposed to evolve bluewards in the 
Hertzsprung-Russell diagram (HRD), and several studies have suggested that they could 
be the progenitors of lower luminosity ($\log$~L/L$_{\odot} \le 5.8$) 
B[e]SGs \citep{Kastner10, Muratore, Aret}.  

On the other hand, B[e]SGs share their location in the HRD with LBVs, 
suggesting that an evolutionary connection between 
these two classes of stars might exist as well. Predicted blueward evolution
 of lower luminosity LBVs suggest they are in a post-YHG (or post-red 
supergiant) phase \citep{HD, Meynet11}. Moreover, S\,65 has 
recently been found to be in a pre-LBV phase \citep[e.g.,][]{Kraus10}.  
Hence the blueward evolution could proceed from YHG through B[e]SG to LBV.
Still, the evolutionary connection in the uppermost part of the HRD
($\log$~L/L$_{\odot} \ge 5.8$) where LBVs and B[e]SGs are also found to coexist
is not well understood, and thus the evolutionary link between 
these most massive stars is yet unknown.  \citet{Morris96} suggested 
that the spectral morphology of S\,18 and two known 
LBV stars, AG Car and P Cygni,
are similar, implying S\,18 is an LBV candidate and providing evidence of 
an evolutionary connection between B[e]SGs and LBVs.  

The results presented in this Letter provide further evidence for the connection between these ambiguous
transition objects.  In addition to the observed variability of CO emission in both S\,65 and HR Car, 
both stars are rotating near critical rotation \citep[valid for HR Car at visual minimum phase; ][]{Groh,Kraus10}, 
and are both located in a similar position on the HRD near to the LBV instability strip determined 
by \citet{Groh}.  Based on the observed properties of S\,65, we assert that 
this star is in a pre-LBV state due to the multitude of shared characteristics, and 
lack of an eruptive event, such as S-Dor-type variability.  
If we presume that all B[e]SGs eventually evolve into LBVs,
what remains particularly puzzling is the timing and physical conditions at which the 
transition from one class to the other occurs, and what role critical stellar rotation plays in influencing 
the transformation.

\section{Conclusion}

Although we have mentioned here only two stars straddling the line between B[e]SG and LBV, it is 
quite possible that a larger number of these stars are evolving toward, or currently in a pre-LBV state (Oksala et al., in preparation).  
Close to critical stellar rotation may support an additional link between these objects.  
As one of the major defining characteristics of LBV stars, the spectral variability of S\,65 and S\,18 suggests concrete evidence of 
their evolutionary connection.  Whether this remains true for all B[e]SGs cannot be determined with the current collection
of observations.  Further monitoring of the spectral features of these fascinating and enigmatic stars 
is essential to reveal the true nature and kinematics of their disks, as well as disclose their evolutionary paths.


\section*{Acknowledgments}

We thank the anonymous referee for valuable comments which improved the manuscript.
 This research made use of the NASA Astrophysics Data System (ADS).
M.E.O. and M.K. acknowledge financial support from GA\,\v{C}R under 
grant number 209/11/1198. The Astronomical Institute Ond\v{r}ejov 
is supported by the project RVO:67985815.  M.B.F. acknowledges Conselho 
Nacional de Desenvolvimento Cient\'ifico e Tecnol\'ogico (CNPq-Brazil) 
for a post-doctoral grant. L.C. acknowledges financial support from the 
Agencia de Promoci\'on Cient\'{\i}fica y Tecnol\'ogica (BID 1728 OC/AR PICT 885), 
PIP 0300 CONICET, and the Programa de Incentivos G11/109 of the Universidad 
Nacional de La Plata, Argentina.  Financial support for International Cooperation 
of the Czech Republic (M\v{S}MT, 7AMB12AR021) and Argentina (Mincyt-Meys, ARC/11/10) 
is acknowledged.  M.F.M. is a research fellow of the Universidad 
Nacional de La Plata, Argentina.  M. L. A. acknowledges finantial support from MINCYT-CONICYT (project CH/11/03).  
M.C. thanks the financial support from CONICYT, Departamento 
de Relaciones Internacionales ÒPrograma de Cooperaci\'on Cient\'ifica InternacionalÓ CONICYT/MINCYT 2011-656 
and from Centro de Astrof\'isica de Valpara\'iso.

\bsp
\label{lastpage}
\end{document}